\documentclass[a4paper,fleqn]{cas-sc}

\usepackage{natbib}
\usepackage[noadjust]{cite}
\usepackage[center]{caption}
\usepackage{subcaption}
\usepackage{url}
\usepackage{hyperref}
\usepackage{graphicx}

\usepackage{amsmath,algorithm,algpseudocode}

\def\tsc#1{\csdef{#1}{\textsc{\lowercase{#1}}\xspace}}
\tsc{WGM}
\tsc{QE}
\tsc{EP}
\tsc{PMS}
\tsc{BEC}
\tsc{DE}
\begin{document}
\let\WriteBookmarks\relax
\def\floatpagepagefraction{1}
\def\textpagefraction{.001}

\shorttitle{CQA-ChatGPT-Impact}

% \shortauthors{CV Radhakrishnan et~al.}

\title [mode = title]{An exploratory analysis of Community-based Question-Answering Platforms and GPT-3-driven Generative AI: Is it the end of online community-based learning?}                      
% \tnotemark[1,2]

% \tnotetext[1]{This document is the results of the research
%    project funded by the National Science Foundation.}

% \tnotetext[2]{The second title footnote which is a longer text matter
%    to fill through the whole text width and overflow into
%    another line in the footnotes area of the first page.}


% ===
\author[1]{Mohammed Mehedi Hasan}
\cormark[1]
\ead{mehedi.bueteee.23@gmail.com}
\affiliation[1]{organization={Department of Computer Science \& Engineering, Independent University, Bangladesh},
    addressline={Bashundhara R/A}, 
    city={Dhaka},
    postcode={1229}, 
    country={Bangladesh}}

\author[1]{Mahady Hasan}
\ead{mahady@iub.edu.bd}
\author[1]{Mamun Bin Ibne Reaz}
\ead{mamun.reaz@iub.edu.bd}

\author[1]{Jannat Un Nayeem Iqra}
\ead{2030087@iub.edu.bd}

\cortext[cor1]{Corresponding author}

\begin{abstract}
% This template helps you to create a properly formatted \LaTeX\ manuscript.
Context: The advent of Large Language Model-driven tools like ChatGPT offers software engineers an interactive alternative to community question-answering (CQA) platforms like Stack Overflow. While Stack Overflow provides benefits from the accumulated crowd-sourced knowledge, it often suffers from unpleasant comments, reactions, and long waiting times. 

Objective: In this study, we assess the efficacy of ChatGPT in providing solutions to software engineering questions by analyzing its performance specifically against human solutions.

Method: We empirically analyze 2564 Python and JavaScript questions from StackOverflow that were asked between January 2022 and December 2022. We parse the questions and answers from Stack Overflow, then collect the answers to the same questions from ChatGPT through API, and employ four textual and four cognitive metrics to compare the answers generated by ChatGPT with the accepted answers provided by human subject matter experts to find out the potential reasons for which future knowledge seekers may prefer ChatGPT over CQA platforms. We also measure the accuracy of the answers provided by ChatGPT. We also measure user interaction on StackOverflow over the past two years using three metrics to determine how ChatGPT affects it.

Results: Our analysis indicates that ChatGPT's responses are 66\% shorter and share 35\% more words with the questions, showing a 25\% increase in positive sentiment compared to human responses. ChatGPT's answers' accuracy rate is between 71 to 75\%, with a  variation in response characteristics between JavaScript and Python. Additionally, our findings suggest a recent 38\% decrease in comment interactions on Stack Overflow, indicating a shift in community engagement patterns. A supplementary survey with 14 Python and JavaScript professionals validated these findings.

Conclusion: Our research concludes that GPT-3 or other LLMs can greatly help software engineers to get a faster, more concise, and more positive response. However, the  implications of reduced community involvement warrant further investigation.

% \noindent\texttt{\textbackslash begin{abstract}} \dots 
% \texttt{\textbackslash end{abstract}} and
% \verb+\begin{keyword}+ \verb+...+ \verb+\end{keyword}+ 
% which
% contain the abstract and keywords respectively. 

% \noindent Each keyword shall be separated by a \verb+\sep+ command.
\end{abstract}


% \includegraphics{figs/grabs.pdf}

% \begin{highlights}
% \item Research highlights item 1
% \item Research highlights item 2
% \item Research highlights item 3
% \end{highlights}

% % Keywords
% % Each keyword is seperated by \sep
% \begin{keywords}
% quadrupole exciton \sep polariton \sep \WGM \sep \BEC
% \end{keywords}

\maketitle

\section{Introduction}
\label{sec:intro}
The field of software engineering has traditionally been regarded as a discipline that promotes the development of communities. The growth was predominantly dependent on the crowd's collective intelligence and the utilization of this knowledge across various community-driven platforms \citep{abdalkareem2017code}. The proliferation of Community Question and Answer (CQA) platforms centered around software engineering subjects has been apparent in recent decades, accompanied by a rise in the involvement of software engineering experts on these websites \citep{bagheri2016semantic}. Stack Overflow, for example, is the pre-eminent online platform for CQA \citep{acar2016you}. As of Dec 2023, over twenty two million individuals have utilized it, and 69\% of the queries on the website have been resolved out of a total of twenty-four million questions asked so far \footnote{https://pgt.page.link/stack-stats}. This platform is commonly utilized by programmers and developers who post various inquiries encompassing software development, programming languages, frameworks, documentation, debugging, and energy consumption and relevant domain experts provide the solution or answers of those. In general users ask questions on CQA platforms to fulfill their desire to know things and apart from cognitive needs, some other motivations like social integration needs and tension free needs or emotion release also play a important role as motivating factors \citep{choi2016user}. On the other hand, people who participate in discussing the questions, answering those or even discussing the answers are often motivated for helping others, reciprocity, and making an impact are more important than financial gains and organizational pressures \citep{penoyer2018impact}.

\par However, prior research also has demonstrated that the desire to obtain a reputation reward can influence the responses on Stack Overflow. This suggests the need for caution when selecting the most suitable answers from the platform. In addition to these challenges, the use of Stack Overflow code has raised concerns about security implications \citep{fischer2017stack}, potential license violations resulting from blind copy-pasting of code \citep{an2017stack}, possible violations of coding and linter best practices \citep{campos2019mining}, challenges in reproducing the problem \citep{mondal2022reproducibility} and a lack of information regarding the waiting time for an answer \citep{goderie2015eta}.

\par On the other side of the scientific advancement front, we have experienced massive growth in Natural Language Processing research, and Large Language Models (LLM) like GPT-3 \citep{floridi2020gpt}, LLaMa \citep{touvron2023llama} are being opened up for general purpose use cases. These models are trained on massive volumes of text data and can appropriately generate natural-sounding text, answer questions, and perform other linguistic tasks. More recently, GPT-3 and ChatGPT have shown state-of-the-art performance on a wide variety of natural-language tasks, such as translation, answering different questions, writing instructed essays, and even writing code, as they are trained on much larger datasets, i.e., texts from massive web corpora including Stack Overflow and Github \citep{touvron2023llama}. Even for zero-shot learning problems, these LLMs have demonstrated superior performance compared to other similar models \citep{qin2023chatgpt}. In general Machine learning uses cases: zero-shot learning aims to complete unlabeled tasks, solving problems without having similar training data \citep{qin2023chatgpt}. Another category of problems that LLMs aim to solve is Chain of Thought (CoT) problems, where they decompose one complex problem into multiple subproblems and then sequentially solve those subproblems with proper reasoning \citep{suzgun2022challenging}. 
Recent studies have examined the similarities and differences between ChatGPT responses and those of human experts, along with the tool's ability to detect plagiarism. A study comparing ChatGPT-generated scientific abstracts to originals using various techniques showed noteworthy findings \citep{gao2022comparing}. Another research evaluated ChatGPT's performance on USMLE Part 1 and Part 2 exam questions, revealing that the model scored above 60 percent, a passing grade for third-year medical students \citep{gilson2023does}. In the legal education context, ChatGPT achieved approximately a C+ level in actual University of Minnesota Law School exams \citep{choi2023chatgpt}. Additionally, a study comparing ChatGPT with human specialists in natural language processing tasks demonstrated its capabilities \citep{guo2023close}. Contrasting perspectives were noted, with some research highlighting distinct linguistic traits in LLMs' writing compared to human-authored essays \citep{herbold2023large}, while others reported inaccuracies and fictitious references in ChatGPT's responses compared to human-generated content \citep{ariyaratne2023comparison}.
\begin{figure}[ht!]
  \centering

  \begin{subfigure}[b]{0.4\linewidth}
    \includegraphics[width=\linewidth]{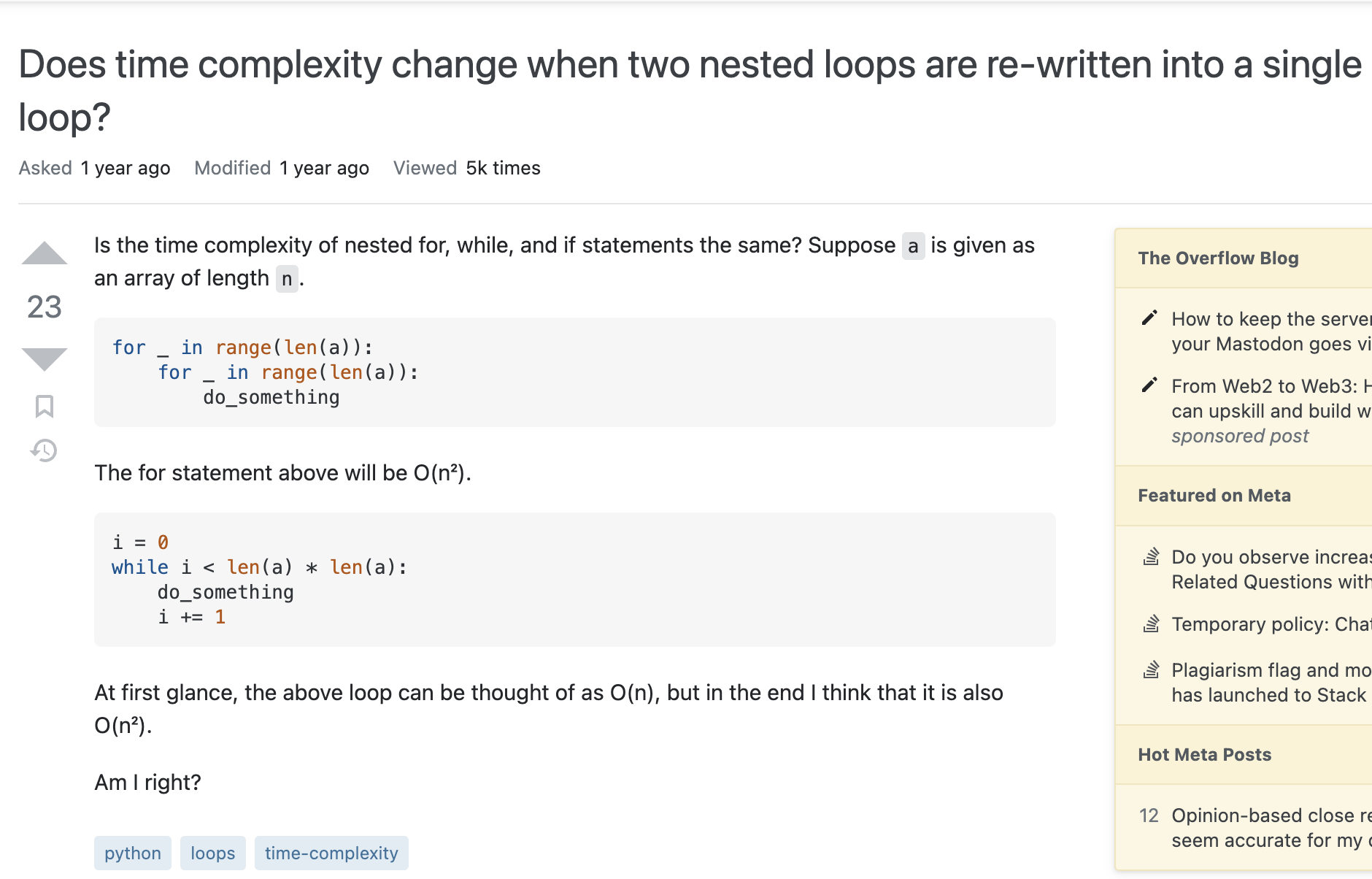}
    \caption{Question Asked in Stack Overflow}
    \label{fig:intro:sub1}
  \end{subfigure}
  \hfill % This adds some space between the subfigures
  \begin{subfigure}[b]{0.4\linewidth}
    \includegraphics[width=\linewidth]{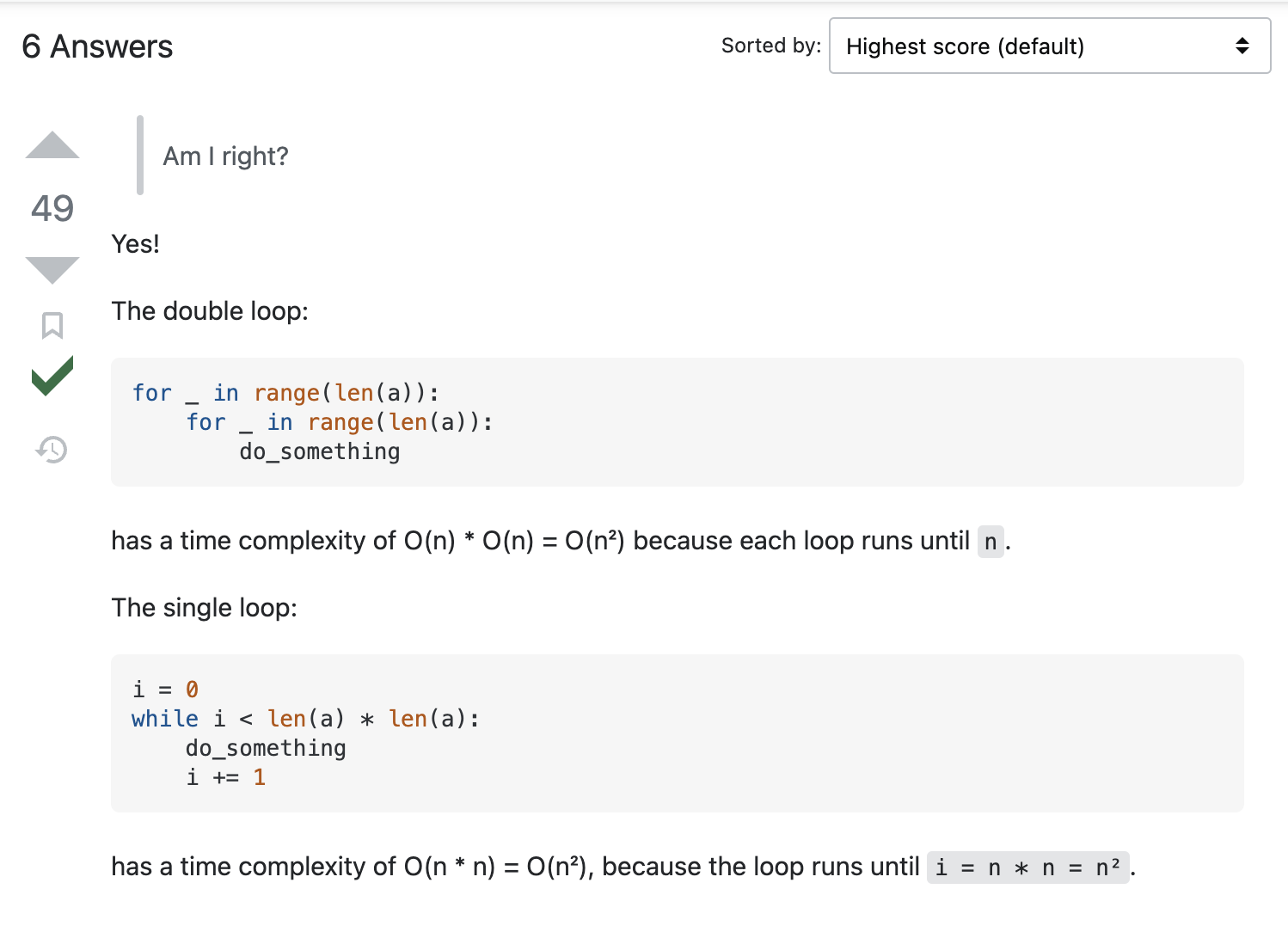}
    \caption{Question Answered in Stack Overflow}
    \label{fig:intro:sub2}
  \end{subfigure}
\hfill
\begin{subfigure}[b]{0.4\linewidth}
    \includegraphics[width=\linewidth]{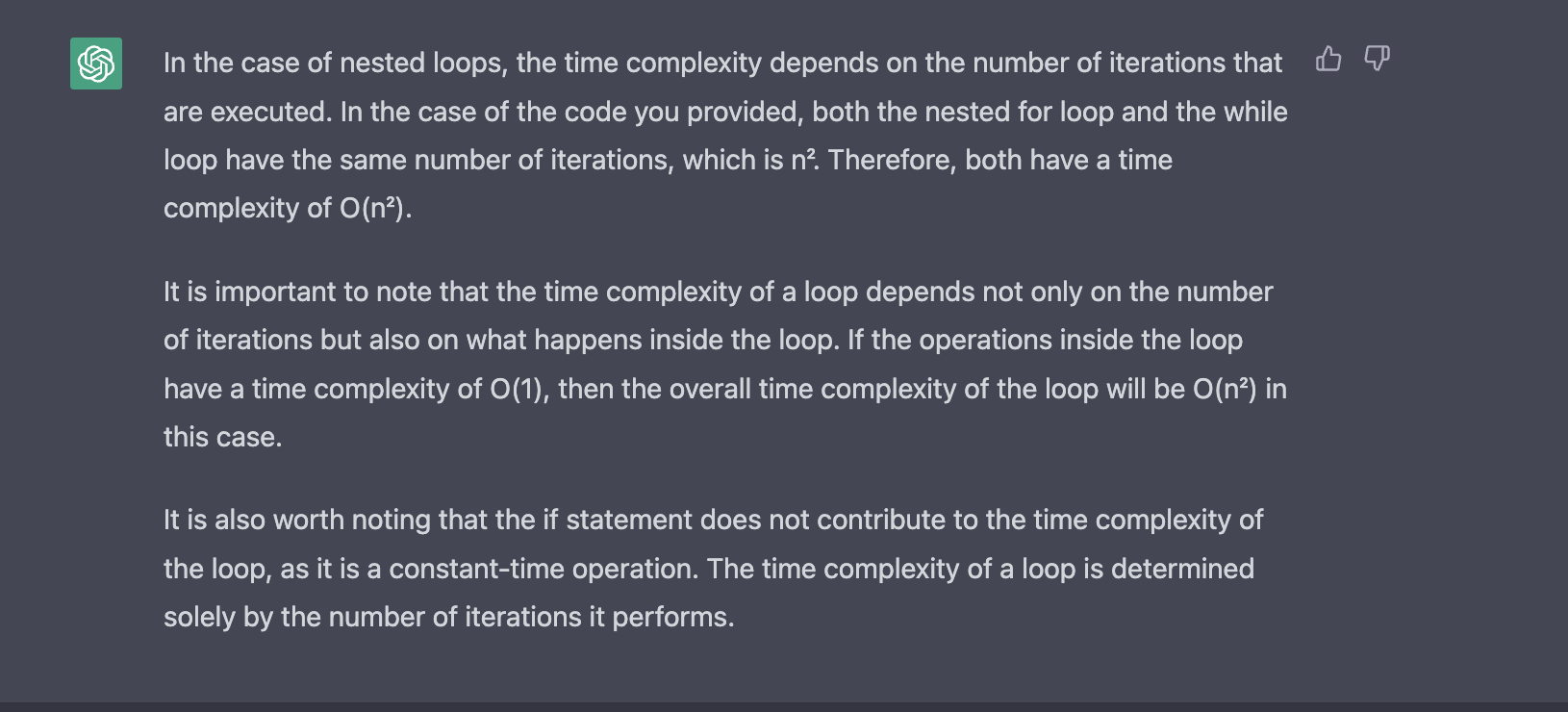}
    \caption{Answer Provided by ChatGPT}
    \label{fig:intro:sub3}
  \end{subfigure}
  \caption{Comparing Question Answer in Stack Overflow and ChatGPT}
  \label{fig:intro}
\end{figure}

\par In software development domain, engineers are primarily motivated by the chance to resolve issues and tackle fresh obstacles to a significant degree \citep{beecham2008motivation}. Furthermore, the capacity to decompose a problem efficiently can assist in resolving intricate and diverse problems \citep{najafi2011multi} of development. These two particular characteristics of software engineering render it an ideal topic for LLM research. ChatGPT and other Language Model Models (LLMs) have significant potential to provide answers to the questions posed on platforms such as Stack Overflow and CQA.

\par To show the real-world motivation behind our research, we demonstrate how GPT-3 models can contribute to answering software engineering-related questions. For example, in Stack Overflow, one user asked a question\footnote{https://pgt.page.link/stack-question} about the time complexity of a nested loop as showed in \ref{fig:intro:sub1}. Another user answered the question with the following clarification and code example as shown in \ref{fig:intro:sub2}. We ask the same question in the ChatGPT UI and get the result shown in \ref{fig:intro:sub3} in the first attempt.

\par The GPT-3 model exhibits a markedly distinct style of explanation and response. A recent study has analyzed answers provided on Stack Overflow and ChatGPT. According to their research, human responses exhibit various variations and frequently stray from the main topic, unlike ChatGPT's responses that consistently focus firmly on the given question \citep{guo2023close}. Nevertheless, it is commonly believed that software engineers acquire more knowledge from their colleagues who work in the same field \citep{murphy2011peer}. Consequently, there is ongoing research on how they will perceive and adopt the responses given by language model systems (LLMs). A recent study by \citep{kabir2023answers} indicates that ChatGPT's responses occasionally need to be revised. 

\par However, fundamental cause for that preference and its consequences on CQA platforms are still being investigated. Individuals may still favor ChatGPT for its other desirable attributes. For example, in Stack Oveflow apart from textual features like length of the code in the answer, reputation of users, similarity of the text between questions and answers, and the time lag between questions and answers, readability have been found as important to make an answer accepted \citep{omondiagbe2019features, zoratto2023study}. As ChatGPT offers near real time interactive question answering, we assume that if it can perform well in other metrics users might prefer ChatGPT over Stack Overflow. In this study, we aim to investigate the following research questions in order to explore those phenomena:

\par\label{rq-1} \textbf{RQ-1:} \textit{Are GPT-3 model provided answers related to software engineering linguistically better than the accepted human answers?}
\par \label{rq-2} \textbf{RQ-2:} \textit{Are GPT-3 model provided answers related to software engineering cognitively better than the accepted human answers?}
\par \label{rq-3} \textbf{RQ-3:} \textit{Are GPT-3 model provided answers related to software engineering has better accuracy than the accepted human answers?}
\par \label{rq-4} \textbf{RQ-4:} \textit{How do human domain experts accept the answers provided by the GPT-3 model?}
\par \label{rq-5} \textbf{RQ-5:} \textit{Is there any change in the usage behavior of Stack Overflow since the public release of GPT-3?}

\section{Methodology}
\par We detail our approach, depicted in Figure 4, in this section.

\begin{figure*}[ht]
    \centering
    \includegraphics[width=0.95\textwidth]{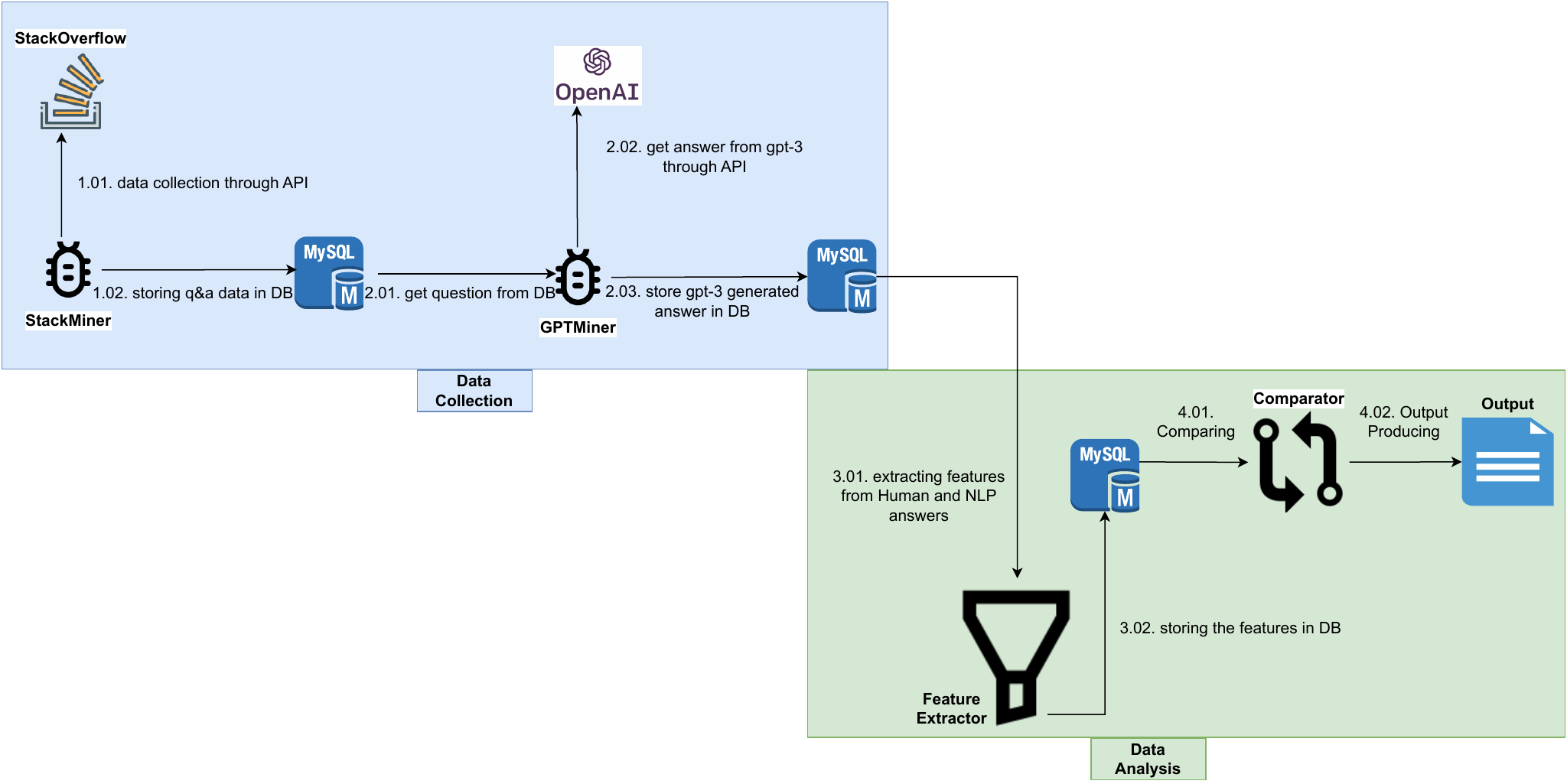}
    \caption{Research Methodology}
    \label{fig:img1}
\end{figure*}

\subsection{Base Question Answer Set Collection}

\par We gather questions from Stack Overflow. To do this, we looked at the most widely used programming languages. We found that the top 5 programming languages in 2022, according to the Stack Overflow developer survey, are typescript, JavaScript, HTML/CSS, SQL, and Python. \footnote{https://pgt.page.link/stack-survey}. Hence, our two programming languages of interest are JavaScript and Python.

\par We use Stack Exchange's RESTful API\citep{richardson2013restful} to collect Stack Overflow questions and question-related metadata\footnote{https://pgt.page.link/question-api}. To collect answer-related metadata, we employ a similar API for answers\footnote{https://pgt.page.link/answer-api}.

\begin{algorithm}[ht]
\caption{StackMiner Algorithm}
\label{Algorithm:01}
\begin{algorithmic}[1]
\Function{fetchQuestion}{$pageNo$, $tag$} 
    \State \Call{questionApi}{$pageNo$, $tag$}
    \State \Return $questions$
\EndFunction
\Statex
\Function{fetchAnswer}{$answerId$} 
    \State \Call{answerApi}{$answerId$}
    \State \Return $answerMeta$
\EndFunction
\Statex
\Function{storeInDb}{$data$} 
    \State $MySQL \gets data$
\EndFunction
\Statex

\While{$question Creation Date <> time Range$}
% \State {$question$ \gets \texttt{$fetchQuestion(pageNo, tag)$}}
\State $question \gets \texttt{fetchQuestion(pageNo, tag)}$
\par
\For{Each question $q$}
    \If{$isAccepted \in q$}
    \State $answerId \gets $q$.id$
    \State $answer \gets \texttt{fetchAnswer(answerId)}$
    \If{$code \in answer$}
    \State $storeInDb \gets answer$
    \EndIf
    \EndIf
\EndFor
\EndWhile
\end{algorithmic}[1]
\end{algorithm}

In both cases, we use API version 2.2. Similar to prior studies\citep{camposmining}, we utilize a few filtering parameters to choose the questions and responses. Our filters were:

\begin{itemize}
    \item Question must have a JavaScript, Python, or typescript tag.
    \item Question must be published between 1st Jan-2022 and 31st Dec-2022.
    \item The Question must have at least one accepted answer.
    \item Answer must have a code snippet.
    
\end{itemize}

We choose the date range between 1st Jan-2022 and 31st Dec-2022 to minimize the probability of these data being used as the training dataset of the ChatGPT model. In later sections, we will choose a model trained with data before January 2022. This is mainly to avoid overfitting the data and ensure the maximum output from ChatGPT's zero-shot and few-shot capabilities. Also, we include the criteria of having code snippets in the question because previous studies \citep{asaduzzaman2013answering} have shown that lack of proper code or program is one of the major causes of unanswered questions.
We extracted metadata from the API response, including the question title, question body, question score, question asker ID, and question date. In addition, we parsed the responses and collected the answer provider ID, the answer content, the answer score, and the answer date. We pull the code snippet that corresponds to the question or answer from the main body of the question or answer. \textit{StackMiner}, a tool built by us, can automatically gather and store all that information in a relational database. In Algorithm 1, we have shown the algorithm used by the stackMiner.

\subsection{Collecting Answer from ChatGPT}
\par To collect the answer from the LLM model, we use OpenAI-driven ChatGPT, similar to some other recent researches\citep{aydin2022openai}. ChatGPT offers different models and engines to use. In our case, we follow the following principles to decide which model to use:

\begin{itemize}
    \item To ensure the base question set was not used as training data for the model, we need a model that was trained before January 1, 2022.
    \item  As questions and answers can sometimes become large, we need a model that offers maximum tokenization.
    \item As the answer can contain code, the model should be able to produce code.
\end{itemize}

\begin{algorithm}[ht]
\caption{GPTMiner Algorithm}
\label{alg:gptAlgo}
\begin{algorithmic}[2]
\Function{fetchQuestionBodyFromDb}{$index$}
    \State ::\Call{DB}{index}
    \State \Return $questionBody$
\EndFunction
\Statex
\Function{fetchGptAnswer}{$answerBody$}
    \State ::\Call{text-davinci-003}{prompt}
    \State \Return $answer$
\EndFunction
\Statex
\Function{storeInDb}{$data$}
    \State $MySQL \gets data$
\EndFunction
\Statex

\State $i \gets 0$
\State $len \gets \texttt{len(dbRecords)}$
\While{$ i \leq len$}
\State $questionBody \gets \texttt{fetchQuestionBodyFromDb(i)}$
\State $gptAnswer \gets \texttt{fetchGptAnswer(questionBody)}$
\If{$\texttt{len(}gptAnswer\texttt{)} \geq 20$}
    \State $storeInDb \gets gptAnswer$
    \EndIf

\EndWhile
\end{algorithmic}[2]
\end{algorithm}

Based on the above criteria, we chose to use GPT-3 models. These models are famous for their capability of understanding and generating natural language texts\citep{dale2021gpt}. Among all the engines of GPT-3, we use text-davinci-003 engine\footnote{https://pgt.page.link/openai-models}. At the time of this research, the text-davinci-003 engine is trained with data up to June 2021 and can process 4000 tokens for each question-answer pair. It is also one of the most capable engines in the GPT-3 model.

\par To collect data using GPT-3, we use the OpenAI Python package\footnote{https://pgt.page.link/opneai-cookbook}, which the OpenAI Foundation has open-sourced. We use the questions we collected from Stack Overflow as the prompt. The API response of OpenAI can give multiple probable answers for each prompt. We used the first response to each prompt. As the text-davinci-003 engine can support a maximum of 4000 tokens, we adjusted the answer token between 512 and 2048, depending on the length of the question. We store the responses to each question in a database. We automate this process with our tool, named \textit{GPTMiner}. In Algorithm 2, we have shown the algorithm we used for GPTMiner.

\subsection{Extracting Metrics}
We obtain separate metrics from the responses from Stack Overflow and ChatGPT. In order to address our research question \ref{rq-1}, we employ four textual metrics such as the number of words in the answer, the length of code in the answer, the similarity between the code in the question and answer, and the overlap of words between the question and answer. Prior research \citep{bian2008finding, fu2019quality} has demonstrated that textual attributes such as answer length and the ratio of question length to answer length influence the acceptability of answers on online question-answer platforms such as Stack Overflow. Additional research \citep{surdeanu2008learning} suggests that a strong correlation between the question and answer often indicates a superior response. The importance of measuring the overlap of words between a question and its corresponding answer has been acknowledged in previous research \citep{burghardt2017myopia}. We utilize the renowned nltk package of Python \footnote{https://www.nltk.org/} to quantify the number of words, sentences, syllables, and stop words.

To answer our \ref{rq-2}, we use some well-known readability metrics, as different readability metrics also play a vital role in the perceived quality of the answers\citep{le2019assessing}. We chose the Automated Readability Index (ARI) \citep{senter1967automated} and the Flesch Reading Ease Score (FRE) \citep{farr1951simplification}, two of the oldest but still highly used metrics. We derive those metrics for Stack Overflow questions, human-provided accepted answers to those questions, and GPT-3 model-rendered answers.

\begin{table}[ht]
\begin{flushleft}
\begin{tabular}{||c c c||} 
 \hline
 \textbf{Name} & \textbf{Type} & \textbf{Equation} \\ 
 \hline\hline
Word Count & Textual (RQ-1) & \eqref{eq-wc} \\ 
 \hline
Code Length & Textual (RQ-1) & \eqref{eq-cl} \\
 \hline
Code Similarity & Textual (RQ-1) & \eqref{eq-sim}\\
 \hline
Word share ratio & Textual (RQ-1) & \eqref{eq-ws} \\
 \hline
Automated Readability Index (ARI) & Cognitive (RQ-2) & \eqref{eq-ari} \\
 \hline
Flesch Reading Ease (FRE) & Cognitive (RQ-2) & \eqref{eq-fre} \\
 \hline
Polarity & Cognitive (RQ-2) & \eqref{eq-polarity} \\
 \hline
Subjectivity & Cognitive (RQ-2) & NA \\
 \hline
\end{tabular}
\caption{Metrics Summary}
\label{table:table-0}
\end{flushleft}
\end{table}

During the calculation of FRE and ARI, we eliminate all codes from the answers. In addition, we conduct sentiment analysis on the responses generated by both the human and GPT-3 models. We assess the polarity and subjectivity of each response using the widely-used TextBlob package \footnote{https://textblob.readthedocs.io/en/dev/install.html}. Prior research has demonstrated \citep{gujjar2021sentiment, bonta2019comprehensive} that TextBlob exhibits a combination of user-friendliness and high functionality, achieving an F1 score exceeding 70\% and an accuracy rate of 74\%. We provide a summary of the metrics, including their type, their relationship with the research question, and their mathematical formulas, in table \ref{table:table-0}. We will provide additional information about those formulas in the following section. 

\subsubsection{ Metrics Definition}

\textit{\textbf{Word Count}}

We consider the effective word count of a text as below:

\begin{equation} 
\label{eq-wc}
WC = \textit{NumOfWords(}text\textit{)} - \textit{NumOfStopWords(}text\textit{)} \end{equation}

\textit{\textbf{Code Length}}

We consider the effective code length of a codeblock as below:

\begin{equation} 
\label{eq-cl}
CL = \textit{NumOfWords(}code\textit{)} 
\end{equation}

\textit{\textbf{Code Similarity}}

To determine the degree of similarity between questions, human replies, and the outputs of the GPT-3 model, we apply cosine similarity. The similarity between two text samples, Text1 and Text2, can be stated mathematically as:

\begin{gather}
\label{eq-sim}
SIM = \frac{A\cdot{B}}{|A|\cdot{|B|}}
\end{gather}

Here, A is the TF-IDF vector of Text1, and B is the TF-IDF vector of Text2. We calculate TF-IDF vectors using Python's sklearn package\footnote{https://pgt.page.link/sklearn} where they are derived from Term Frequency and Inverse Document Frequency. Mathematically, those can be written as follows:
\begin{gather}
TF = \frac{\textit{Count of a term in a document}}{\textit{Count of all terms in the document}} \\
IDF = \ln{\frac{\textit{Total count of documents}}{\textit{Count of documents with the term in it}}}
\end{gather}

\textit{\textbf{Word Share}}

We calculate word share by following simple formula:

\begin{equation}
\label{eq-ws}
    WS = \frac{\text{\textit{len}(Question Words}\bigcap \text{Answer Words)}}{\text{\textit{len}(Answer Words)}}
\end{equation}

\textit{\textbf{Automated Readability Index (ARI)}}
 
 The automated readability index (ARI) is a tool used to evaluate the readability of written English. It provides a rough estimate of the US school grade level at which the text must be read for comprehension. A text with a score of 10 is understandable by a 9th grader, whereas a text with a score of 14 or higher requires at least a college education to fully comprehend \footnote{https://en.wikipedia.org/wiki/Automated\_readability\_index}. ARI focuses on the words and character count of the text, and we can derive ARI with the following formula:

\begin{gather}
\label{eq-ari}
    ARI = 4.71 * \frac{\textit{total characters}}{\textit{total words}} + 0.5 * \frac{\textit{total words}}{\textit{total senteces}} -21.43
\end{gather}

\textit{\textbf{ Flesch Reading Ease (FRE)}}

On the other hand, the Flesch Reading Ease (FRE) score also indicates the readability of a certain text. Instead of emphasizing individual characters, FRE emphasizes syllables. When grading a passage, a score of 60–70 indicates that a ninth-grader will have no trouble understanding it, while a score of 30 or lower indicates that a minimum of a college education is required \footnote{https://simple.wikipedia.org/wiki/Flesch\_Reading\_Ease}. The formula to derive FRE is:

\begin{gather}
\label{eq-fre}
    FRE = 206.835 - 1.015  * \frac{\textit{total words}}{\textit{total sentences}} -84.6 \frac{\textit{total syllables}}{\textit{total words}}
\end{gather}

\textit{\textbf{ Polarity}}

 The polarity of the text is a number between -1 and 1, with -1 indicating very negative sentiment and +1 being positive. Mathematical analysis of the text's positive and negative word frequencies yields the polarity scores. A high-level mathematical equation can be:

 \begin{gather}
\label{eq-polarity}
Polarity = \frac{PW - NW}{TW}
\end{gather}

 Where \textit{PW} is the number of positive words, \textit{NW} is the number of negative words and \textit{TW} is the number of total words in a text. When measuring this,  TextBlob package removes the stop words and other non-relevant words from the text.

 \textit{\textbf{ Subjectivity}}
 
 Opinions and assessments are examples of subjective output, which ranges from [0,1]. The subjectivity scores is calculated using a mathematical equation that considers the  degree of opinionated statements. 

 \subsection{Qualitative Analysis for measuring Accuracy}
 To find the answer of \ref{rq-3} we need to quantify the precision of the responses which is a complex task. Sometimes, a question can have multiple correct answers depending on the specific circumstances. We utilize manual analysis and open coding techniques described by \citep{saldana2021coding} at this location. In order to conduct a manual analysis, we selected a random sample of 100 questions from the Python dataset and another 100 questions from the javascript dataset. The primary author, possessing nine years of expertise in Javascript and five years of proficiency in Python, manually scrutinized the 200 questions. Given the dataset's diverse range of topics, we dedicated 140 hours to thoroughly analyzing and comprehending the entire dataset. Due to the implementation of random sampling on the entire dataset, there are specific inquiries that the GPT-3 model cannot respond to. The initial author assigned two labels to each question: \textit{is\_answered\_by\_chatgpt} and \textit{is\_chatgpt\_answer\_correct}. The initial author required an additional 80 hours to complete this labeling task. In order to determine the accuracy of an answer, the primary author utilizes the following procedure, which has been unanimously approved by the research team:

 \begin{itemize}
    \item If the answer of ChatGPT proposes exactly same solution as the accepted human answer, it is a correct answer
    \item If the answer from ChattGPT is not matching with the accepted answer but pass any of the following three checks, it is a correct answer:
     \begin{itemize}
         \item ChatGPT's answer is matching with other non-accepted answers which has been upvoted by users as correct answer
         \item Manual reproduction of the ChatGPT provided code snippets is producing the desired result
         \item In worst case, if both of the above is not successful. but the solution proposal is supported by documentation of the software or packages
     \end{itemize}
    \item If non of the correctness criteria is matched or ChatGPT could not answer that question we consider that as an incorrect answer
\end{itemize}

 The categorization process is vulnerable to rater bias since only the first author identifies these categories. We allocated an additional rater, the software engineering research assistant, to address this constraint. The second author possesses over three years of expertise in Software Engineering research. Subsequently, the initial author provides an introduction and imparts knowledge regarding the dataset. The second evaluator autonomously reviews each question, the responses from humans, and the responses from ChatGPT. Subsequently, the second rater assign identical labels to the ChatGPT responses after spending around 200 hours. Subsequently, we calculate a Cohen's Kappa coefficient \citep{saldana2021coding} of 0.86 between the primary and secondary author, signifying a nearly flawless agreement.
 
\subsection{Collecting Domain Expert Opinion}
\par We employed a random selection process to choose two questions to gather  human experts' viewpoint, explicitly focusing on the JavaScript and Python tags. In order to mitigate the influence of extraneous factors such as explanations and code examples, we opted to conduct a survey consisting of only two questions. Having all participants evaluate the identical question can enhance the comparability between ChatGPT and human responses. Alternatively, numerous factors related to questioning can weaken the comparison. We conducted two surveys to gather data from the domain experts. Each survey includes the original question, an accepted answer from a human Stack Overflow user, and an answer generated by GPT-3. We solicited input from domain experts to assess their stance on various statements using a 5-point Likert scale. A rating of one indicates a strong disagreement with the statement, while a rating of five indicates a firm agreement. During a survey, we presented the response provided by a human as the primary option, followed by the answer generated by the GPT-3 model as the secondary option. In the previous survey, we displayed the data reversed to prevent potential bias based on position. Fourteen software engineering professionals participated in the survey by completing a structured questionnaire. The participants were selected using convenience and purposive sampling methods. Each participant had at least five years of professional experience in their fields and had successfully finished a graduate program in various engineering disciplines at the university level.

\subsection{Collecting Stack Overflow Usage Trend}
We use Stack Exchange Data Explorer Query Interface \footnote{https://data.stackexchange.com/Stack Overflow/query/edit/1787356} to collect usage behavior. We gather three metrics to analyze the behavior: \textit{Monthly Question Count}, \textit{Monthly Comment Count} and \textit{Monthly New User Count}. We take data for last 2 years, e.g from Sep-2021 to Sep-2023. We use the following SQL query to extract the data for each month between the time period:

\begin{verbatim}
count = SELECT COUNT(Id) FROM $tableName$  
WHERE CreationDate > $startDateOfEachMonth$  
AND CreationDate < $startDateOfNextMonth$ 
\end{verbatim}
Here we used three tables from Stack Overflow database named: Posts, Comments and Users.

\subsection{Data Overview}

We initially mined 4153 questions from Stack Overflow, of which 1704 have JavaScript or TS tags and 2499 have Python tags. All of those questions have one accepted answer. Out of 4153 answers, answer providers gave code examples and snippets in 4032 answers. We pass these 4032 questions on to GPTMiner. Some of the answers to the GPT-3 model need to be more relevant, and some cannot be answered. So we put a threshold of at least 20 words in the GPT-3 model's answer. From there, we get 2693 answers. After some other cleanup due to unreadable answers, we finally locked in 2564 questions and answers as our base dataset.
\begin{table}[ht]
\begin{flushleft}
\begin{center}
\begin{tabular}{||c c||} 
 \hline
 \textbf{Heading} & \textbf{No of Data} \\ [0.5ex] 
 \hline\hline
 Initially Collected Questions & 4153 \\ 
 \hline
 JavaScript/Typescript Tags & 1704 \\
 \hline
 Python Tag  & 2449 \\
 \hline
 Human Answer having code snippets & 4032 \\
 \hline
 GPT-3 Model's answer length $>$ 20 & 2693 \\ 
 \hline
 Final Data Set & 2564 \\
 \hline
 GPT-3 Model's answer having code snippet in final dataset & 1575 \\ [1ex] 
 \hline
\end{tabular}
\end{center}
\caption{Data Summary}
\label{table:table-1}
\end{flushleft}
\end{table}

Each piece of data has attributes such as the title of the question, the body of the question, code snippets provided in the question, the body of the answer, code snippets shared in the answer, the body of the GPT-3 answer, and the code snippets included in the GPT-3 answer. All that information is stored in a SQL table named base-data. A summary of this data has been provided in table \ref{table:table-1}.

From this table data, we apply the metric derivation formulas described in the feature extraction section, derive the metrics, and finally store those in a table called metrics. This table has attributes like question word count, answer word count, GPT-3 model's answer word count, question code length, answer code length, GPT-3 model's code length, FRE, ARI of both human answers, and GPT-3 model's answers, word share with a question for both types of answers, polarity, and subjectivity of both types of answers.

\section{Result}
\par We will discuss the output of our data analysis in this section. In software development, apart from mean and median, we often use 95 percentiles and 99 percentiles to see the coverage value of a metric. We will use the mean, 95 percentile, and 99 percentile of different metrics to discuss the result.

\subsection{\textbf{Answer to RQ1:} Textual Feature Analysis}
Word count has been considered a critical textual metric in previous literature to determine the effectiveness of an answer in a stack overflow. We determine the word count of questions, accepted human answers, and answers provided by the GPT-3 model. For word count, we consider only the sample data with at least 20 words in the answer of the GPT-3 model. We imposed this condition to eliminate the data where GPT-3 could not provide valid answers.

\begin{figure}
  \centering
  \includegraphics[width=3in]{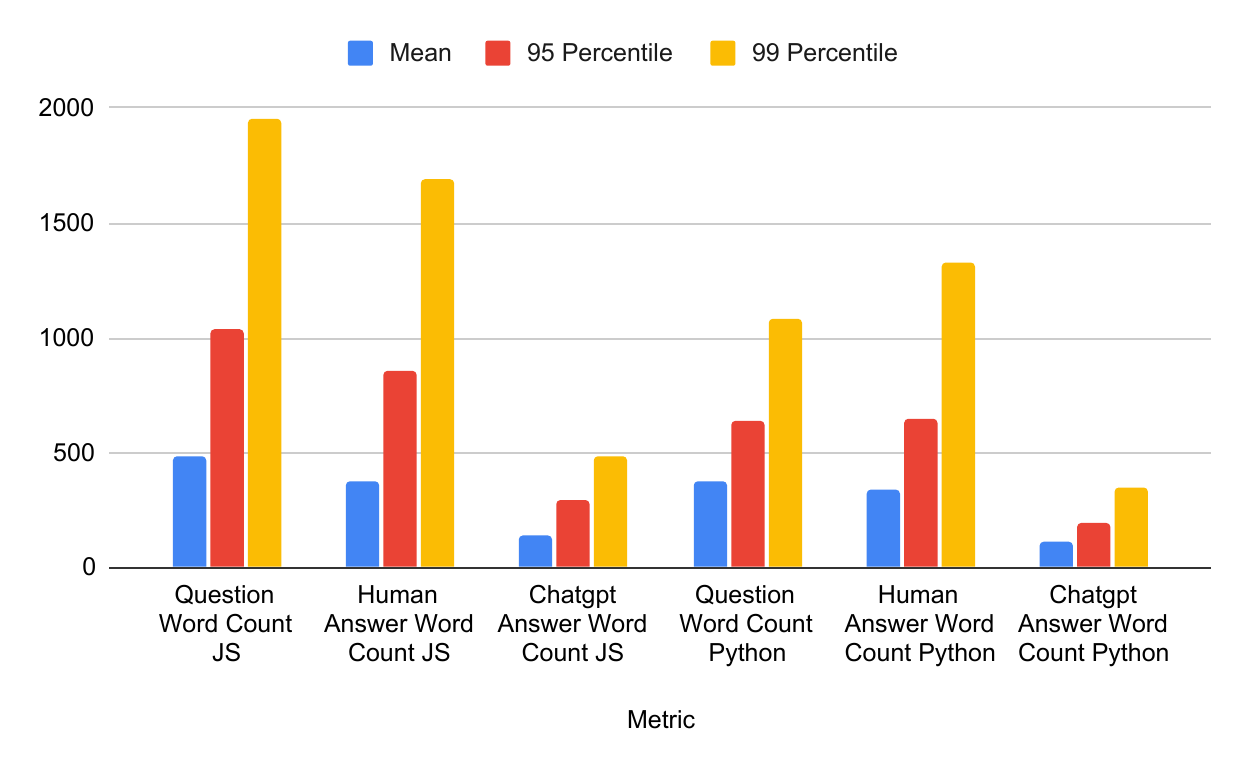}
  \caption{Word Count Comparison}
  \label{fig-wc}
\end{figure}

The mean, 95 percentile, and 99 percentile word counts of human-provided answers for JavaScript-tagged questions are 375.56, 853, and 1690; for Python-tagged questions, they are 340.12, 648, and 1325, whereas the same metrics for the GPT-3 model answer for the first tag are 138.63, 290, and 483; and for the second tag, they are 116.17, 197, and 347. Figure \ref{fig-wc} indicates that even at the 95 percentile level, human answers are almost twice as verbose as the GPT-3 model answer for both JavaScript and Python. However, it is worth noticing that the GPT-3 model provided Python-tagged answers that had 12\% less density of word count concerning human answers than JavaScript answers. We try to find a correlation between the questions in Figure 6, irrespective of the language. The Pearson correlation coefficient between the question word count and the human-provided answer word count is around 0.24, and the same between the question word count and the GPT-3 answer word count is approximately 0.14. This also indicates that GPT-3 models can deduce the question summary from an extensive text and give a more concise output.

\begin{figure}
  \centering
  \includegraphics[width=3in]{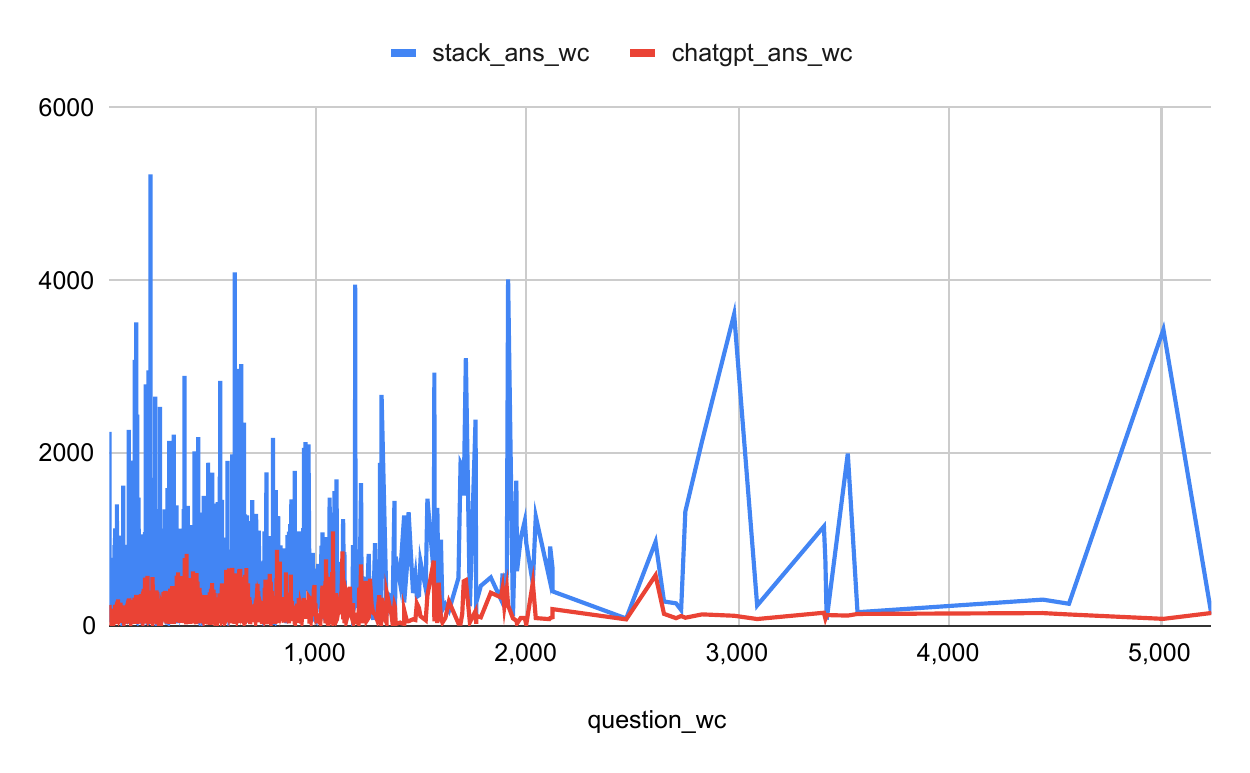}
  \caption{Answer Word Count Distribution w.r.t Question Word Count}
  \label{fig-wc-dist}
\end{figure}

\begin{figure}[ht]
  \centering
  \includegraphics[width=3in]{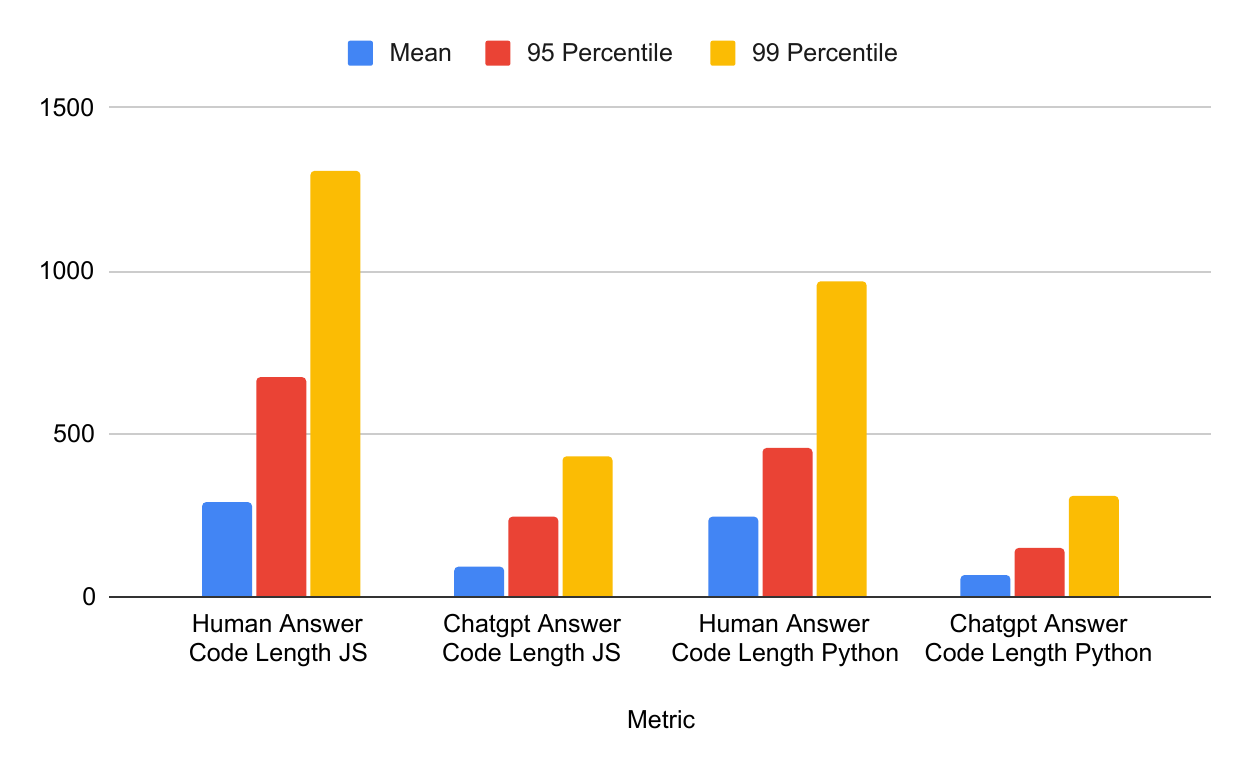}
  \caption{Code Length Comparison}
  \label{fig-codelen}
\end{figure}

In Stack Overflow answers, answer givers often provide code examples for the described problem. We measure the code length of the human-provided answer and the GPT-3 model-provided answer, and in Figure \ref{fig-codelen}, we illustrate the metrics. Apart from the GPT-3 answer word count, we also introduce one more constraint on question code length (min. 25) while analyzing this part to avoid questions where the askers provided no code.

\begin{figure}[ht]
  \centering
  \includegraphics[width=3in]{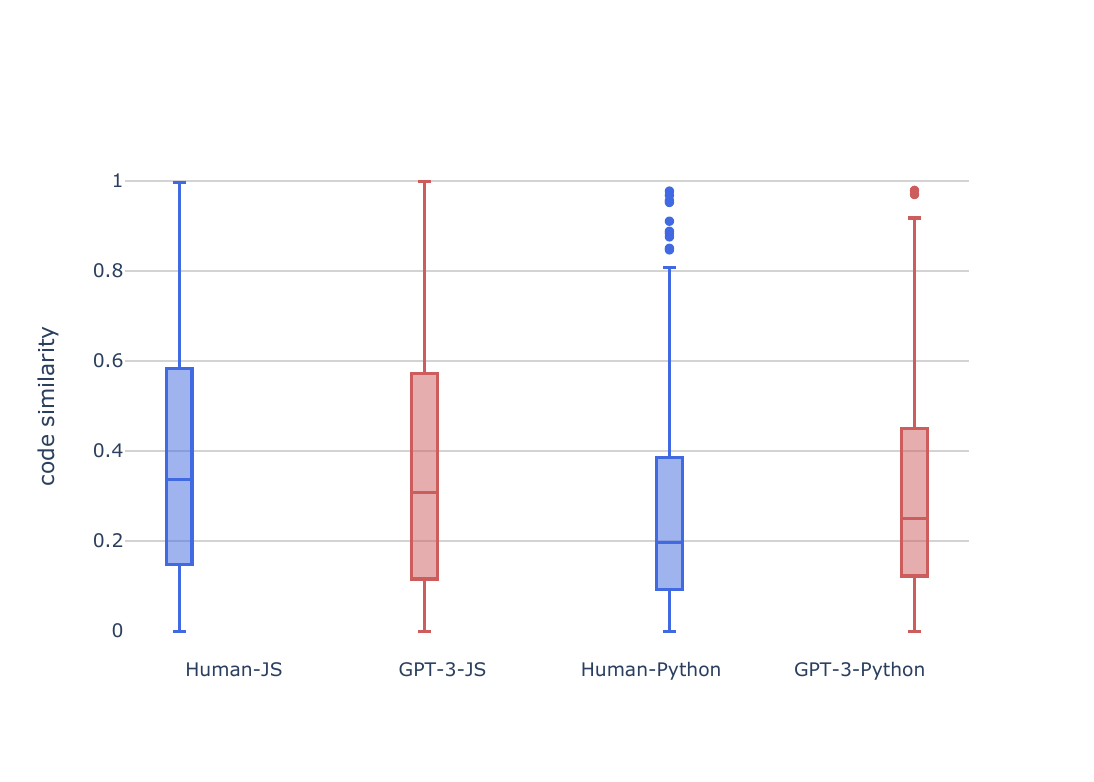}
  \caption{Cosine Similarity of Question code and Answer Code}
  \label{fig-codesim}
\end{figure}

We observe a low volume of code output in the GPT-3 model, similar to the word count. Here, the Pearson correlation coefficient between question code length and human answer code length is 0.34, whereas, with the GPT-3 model, the code length is 0.09. It indicates that, even if the askers write much code in the question, the GPT-3 model may not provide code examples in the answer, and even if it does, the code snippets will be tiny compared to the question code. Even when we examine language or tag level, we see that on the 95 percentile for both Python and JavaScript questions, the GPT-3 model produced almost one-third of the code compared to human-provided code from a volumetric point of view (figure \ref{fig-codelen}). We then examine the similarity between the code that askers provide as a problem and the answer provided by the human vs. GPT-3 model in Figure \ref{fig-codesim}. Here, we can see that human-provided code and GPT-3 model-provided code are closely similar to the code provided by the asker for JavaScript-tagged questions; even human-provided code is ahead in this metric. However, for Python-tagged questions, we see the GPT-3 model's provided codes having a higher similarity with the question code. Note that along with the previous two constraints, we added two more on human-provided code length (min. 25) and the GPT-3 model-provided code length (min. 25) while analyzing the similarity to avoid outliers.

\begin{figure}[ht]
  \centering
  \includegraphics[width=3in]{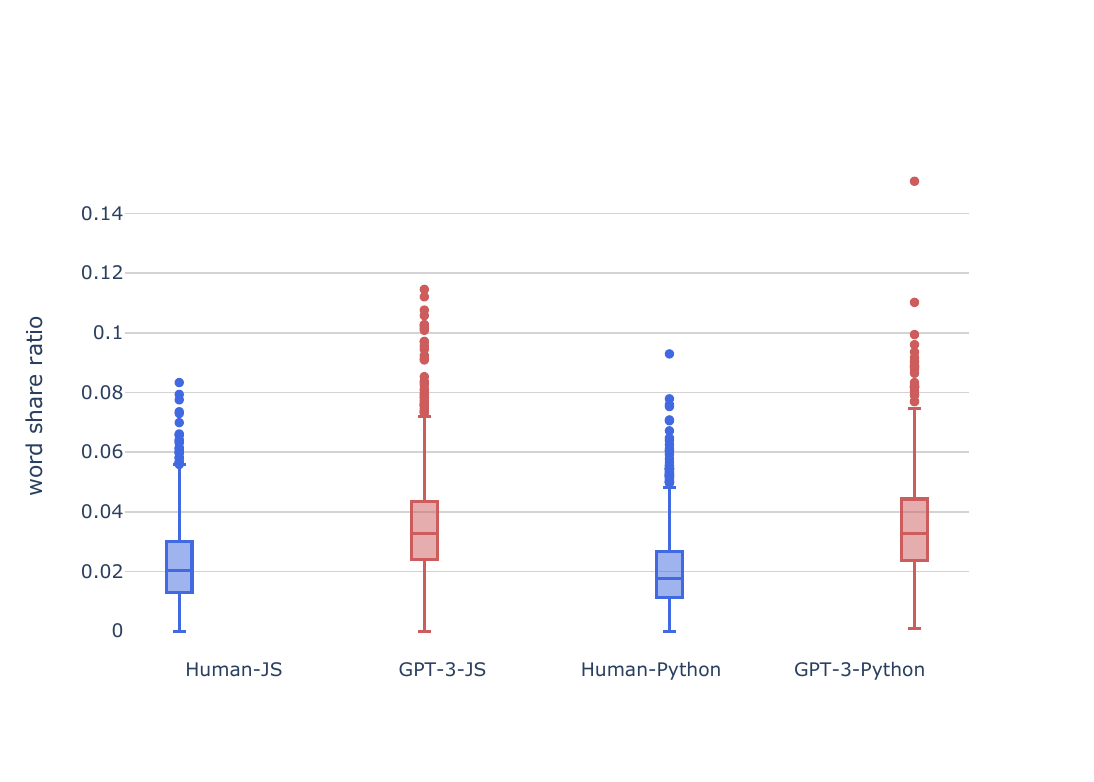}
  \caption{Word Share comparison between Human answer and GPT-3 Model Answer}
  \label{fig-wordshare}
\end{figure}

On the other hand, when we examined the word share ratio between the question and answer, as shown in figure \ref{fig-wordshare}, we observed significantly higher word share between the question and the GPT-3 model provided answer than the human answer in both JavaScript and Python tagged questions. 

Where human answers have a word-share ratio with the original question of around 0.022, 0.048, and 0.061 on the mean, 95 percentile, and 99 percentile, we see on the exact measurement that the GPT-3 model shows the ratio as 0.035, 0.065, and 0.89, which indicates that on the 95 percentile level, GPT-3 responses contain 35.4\% more common words with questions.

\subsection{\textbf{Answer to RQ2:}Cognitive Feature Analysis}
\begin{figure}[ht]
  \centering

  \begin{subfigure}[b]{0.4\linewidth}
    \includegraphics[width=\linewidth]{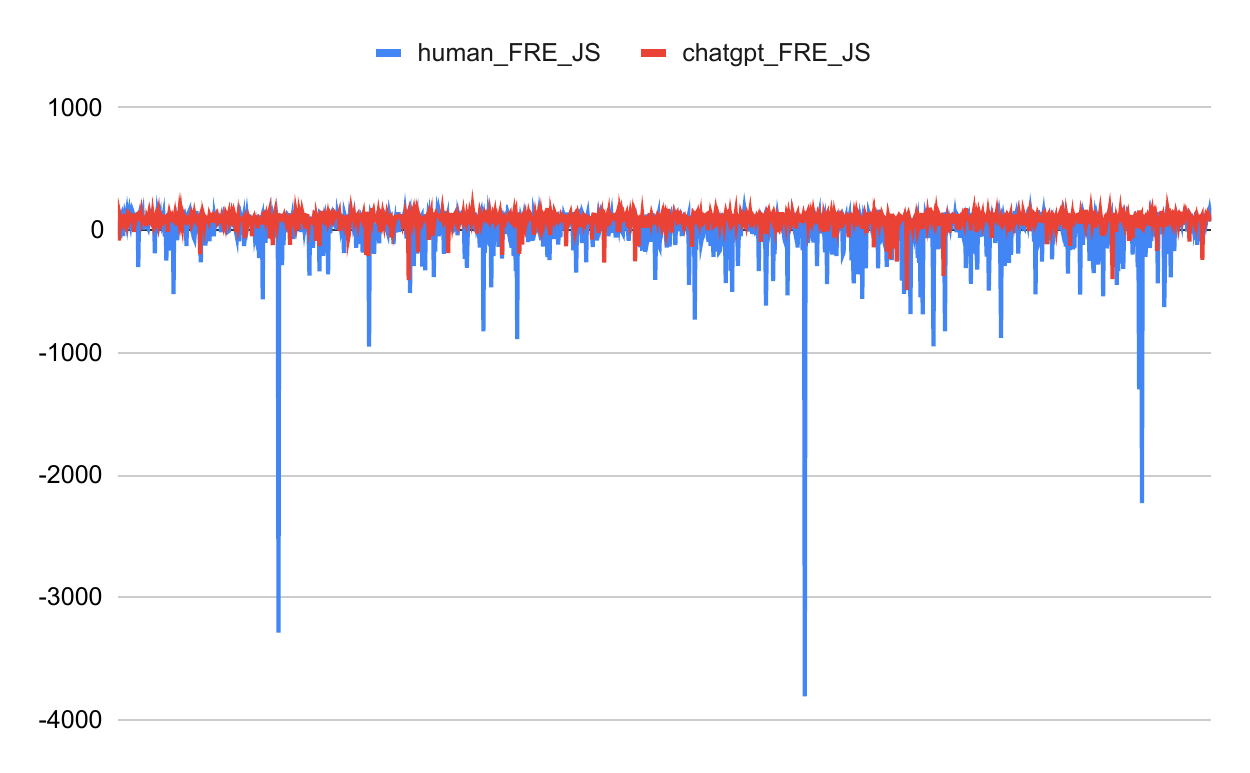}
    \caption{JS FRE Comparison}
    \label{fig:fre:sub1}
  \end{subfigure}
  \hfill % This adds some space between the subfigures
  \begin{subfigure}[b]{0.4\linewidth}
    \includegraphics[width=\linewidth]{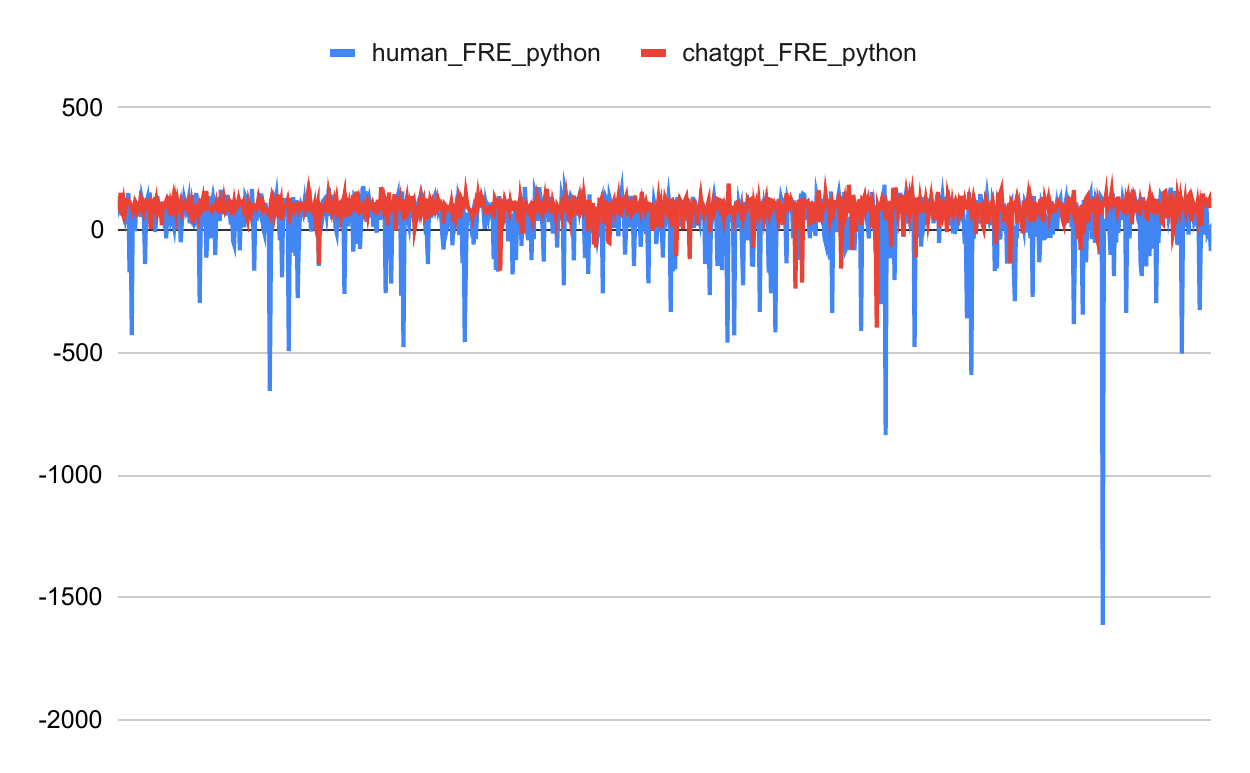}
    \caption{Python FRE Comparison}
    \label{fig:fre:sub2}
  \end{subfigure}

  \caption{FRE comparison between Human answer and GPT-3 Model Answer}
  \label{fig:fre}
\end{figure}

We chose the Flesch Reading Easiness (FRE) index and the Automated Reading Index (ARI) to analyze the cognitive features of the answers. Similar metrics have been used in previous literature also \citep{le2019assessing, burghardt2017myopia}. Lower scores on the FRE indicate that the document is more difficult to read, whereas higher levels indicate better readability. Less than 30 indicates university-level reading comprehension. Readings with scores between 90 and 100 are deemed suitable for a 10-year-old child \citep{squire2014bit}. Figure \ref{fig:fre} demonstrates that GPT-3 model responses have a substantially more consistent and higher FRE than human responses, with a median of 97.88 and a mode of 125.29 compared to 65.75 and a mode of 74.88 for human responses. Interestingly, when we drill down language-wise, we find that GPT-3's FRE mode for JavaScript questions is 67\% higher than human-provided answers. In contrast, for Python questions, the FRE mode of human-provided answers is 7\% higher.

\begin{figure}[ht]
\centering
\begin{subfigure}{.4\linewidth}
  \centering
  \includegraphics[width=\linewidth]{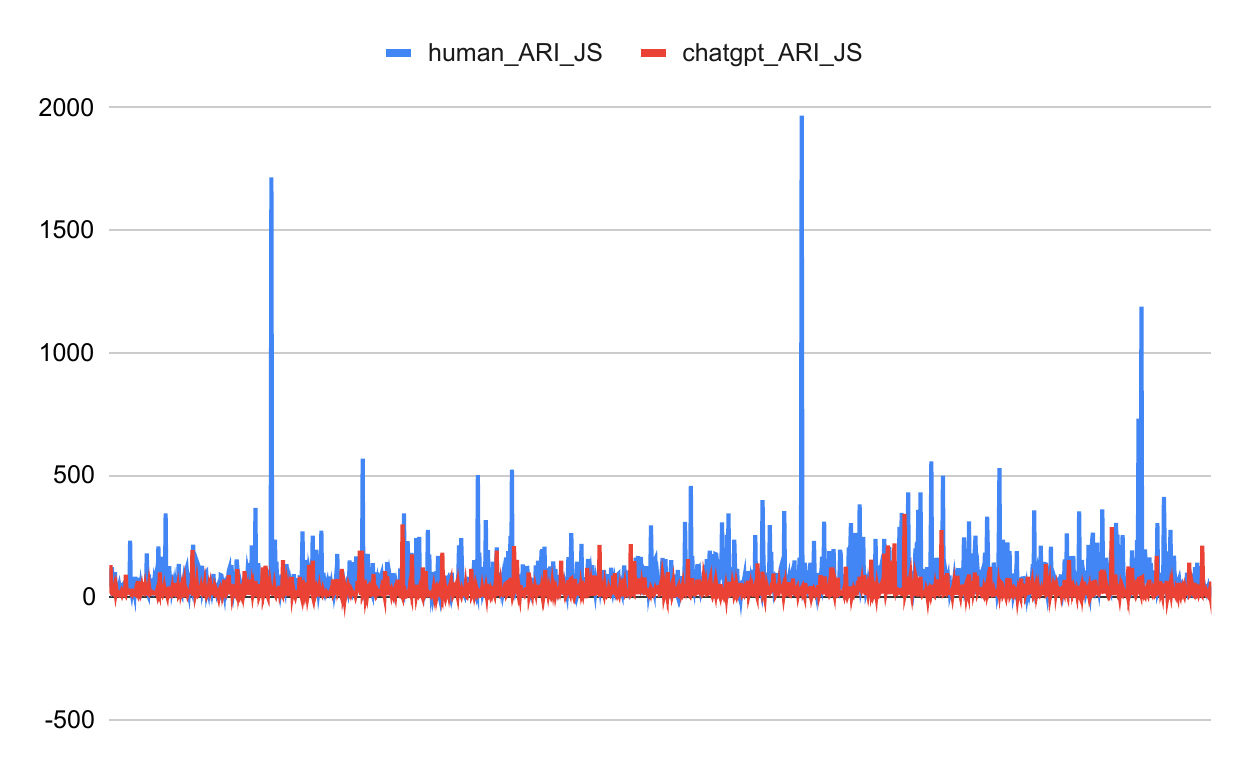}
  \caption{JS ARI Comparison }
  \label{fig:ari:sub1}
\end{subfigure}%
\hfill % This adds some space between the subfigures
\begin{subfigure}{.4\linewidth}
  \centering
  \includegraphics[width=\linewidth]{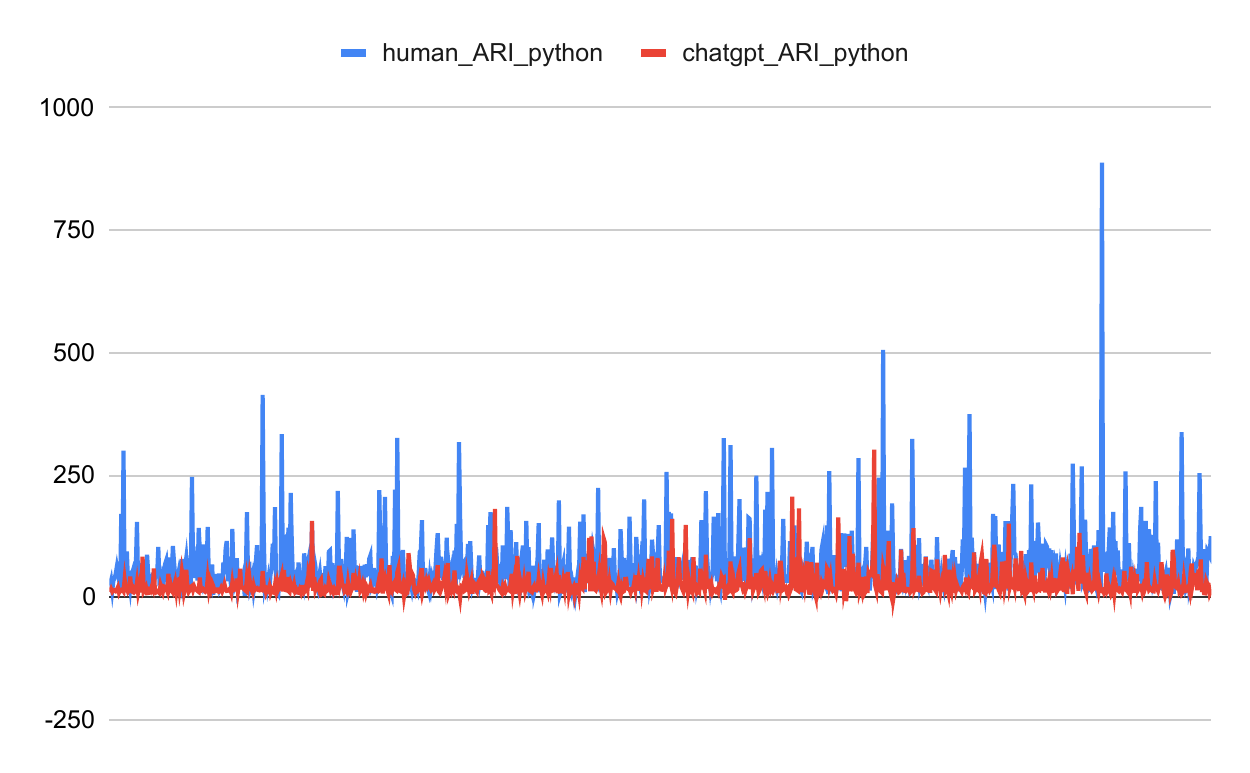}
  \caption{Python ARI Comparison}
  \label{fig:ari:sub2}
\end{subfigure}
\caption{ARI comparison between Human answer and GPT-3 Model Answer}
\label{fig:ari}
\end{figure}

When considering ARI, a score of around figure \ref{fig:ari} indicates that a 10th grader can understand the text, whereas a college student will understand it with a score greater than 14. This comparison shows that human and GPT-3 model answers are beyond the level of understanding of a college student. However, human-provided answers have many outliers, which should be harder for users to understand. Here, language-wise, we also see a similar trend, where the median and mode of the ARI of human-provided JavaScript answers are 54 and 29.6, respectively, and Python answers are 45.2 and 55.9. In contrast, GPT-3 model-provided JavaScript answers have a median and mode ARI around 21.6 and 13.5, respectively, and Python answers have a median and mode ARI around 18.6 and 22.5.

\begin{figure}[ht]
  \centering
  \includegraphics[width=3in]{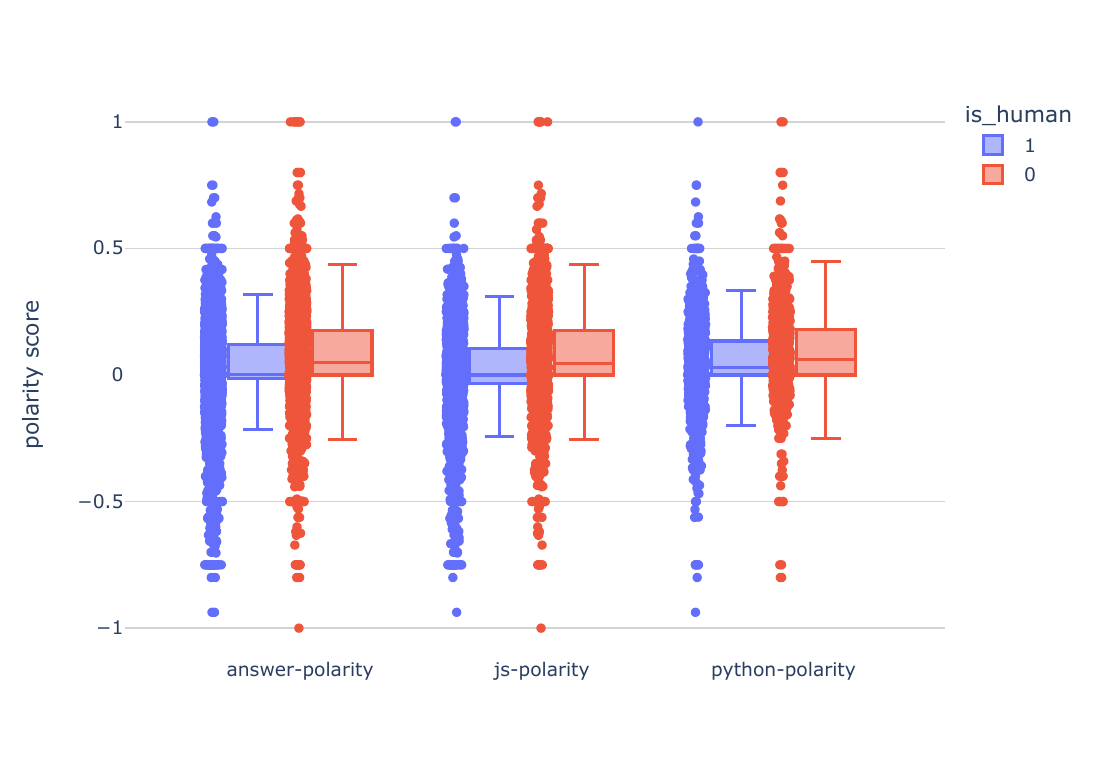}
  \caption{Polarity comparison between Human answer and GPT-3 Model Answer}
  \label{fig-polarity}
\end{figure}

Finally, we can see the polarity distribution in the answers provided by the human and GPT-3 models in Figure \ref{fig-polarity}. Human answers are colored blue, and the GPT-3 model's answers are red. The mean, 95 percentile, and 99 percentile polarity scores of human answers were generally 0.022, 0.32, and 0.5. In contrast, the GPT-3 model's answers were 0.076, 0.4, and 0.6, irrespective of programming language, indicating that on the 95 percentile level, GPT-3 shows 25\%  higher positive sentiment in its response than the human being. One noteworthy observation is that, in general, GPT-3 model answers have more positive sentiment than human answers in both Python and JavaScript-tagged questions. However, Python answers from the GPT-3 model carry more positive sentiment than JavaScript answers.

\begin{figure}[ht]
  \centering
  \includegraphics[width=3in]{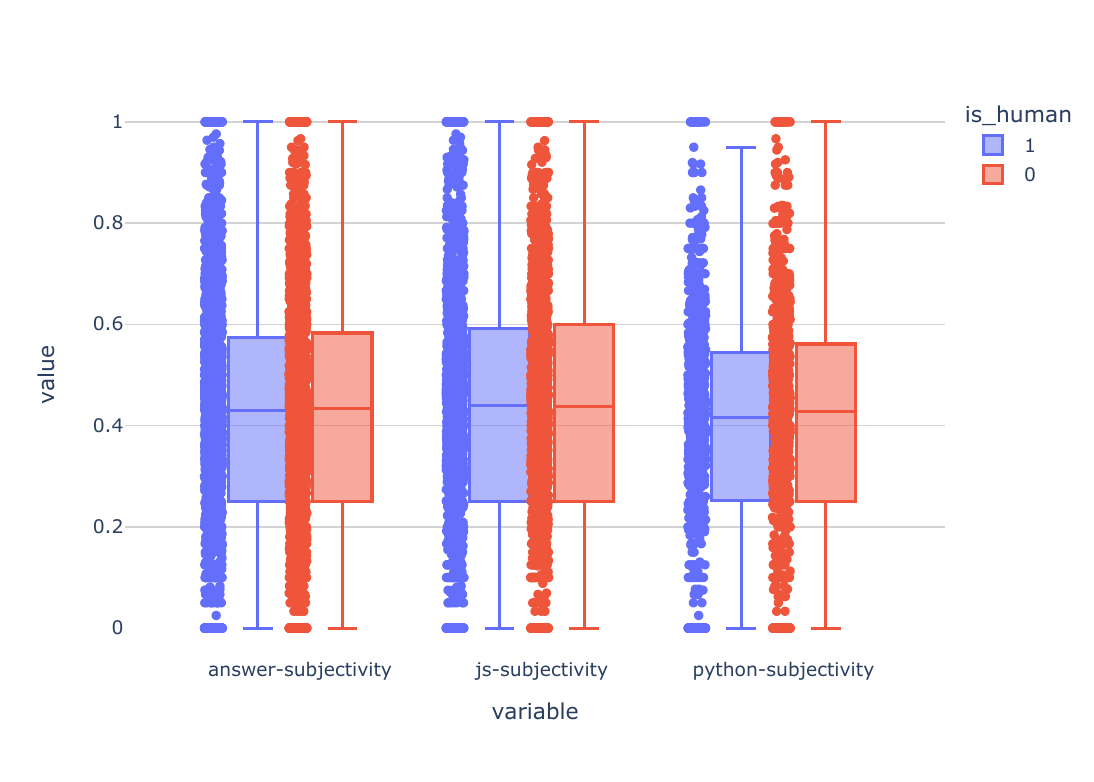}
  \caption{Subjectivity comparison between Human answer and GPT-3 Model Answer}
  \label{fig-subjectivity}
\end{figure}

Correspondingly, when we analyze the distribution of subjectivity for the same samples, we observe no notable distinction between the responses of human and GPT-3 models. Both human and GPT-3 model subjectivity ranges from 0.25 to 0.425, with a median score higher than 0.425. Figure \ref{fig-subjectivity} indicates that both types of answers are more subjective and contain a moderate level of opinion or emotion. It suggests that the text may present some facts or information but will likely contain personal opinions or subjective interpretations.

\subsection{\textbf{Answer to RQ3:} Accuracy Analysis}
Out of 100 randomly sampled Javascript questions, ChatGPT could answer 82 questions. In the first verification version, they rated 71 answers out of 82 as correct. The second rater found 69 answers to be correct. Both raters discussed the conflicting questions and answers, and the second rater fully agreed with the first rater.

Similarly, ChatGPT could answer 95 questions out of 100 for Python questions. Out of those 95 answers, 75 were concluded as correct by some arguments among both raters. Finally, there was 100\% agreement between the raters. Finally, we calculate that on Javascript answers, ChatGPT's accuracy was 71\%, and on Python, it is accuracy was 75\%.

\subsection{\textbf{Answer to RQ4:} Domain Expert Response Analysis}
We collected responses from 14 domain experts who are not authors of this research. Each has a minimum of five years of work experience and has completed a university-level undergraduate degree in Engineering. As mentioned in table \ref{table:table-2}, we request the participants rate their level of agreement with five statements using a Likert scale, then compile their replies. We consider a response greater than 3 to be a favorable agreement with the claim \citep{boone2012analyzing}. 57\% of respondents preferred human answers for metrics like accuracy and example, whereas 65\% preferred GPT-3 model answers for metrics like readability, conciseness, and explanation. The mode, mean, and median of the Likert Scale survey response have been summarized in the table \ref{table:table-2}.

\begin{table}[ht]
\begin{flushleft}
\begin{center}
\begin{tabular}{||c c c c c||} 
 \hline
 \textbf{SL} & \textbf{Question} & \textbf{Mode} & \textbf{Mean} & \textbf{Median} \\ [0.5ex] 
 \hline\hline
1 & Human answer is more accurate & 4 & 3.54 & 4  \\ 
 \hline
2 & Human answer is more easily readable  & 3 & 2.5 & 3 \\
 \hline
3 & Human answer is more concise  & 2 & 2.6 & 3 \\
 \hline
4 & Human answer has better explanation  & 1 & 2.1  & 3\\
 \hline
5 & Human answer has better example  & 4 & 3.1 & 4 \\ [1ex] 
 \hline
\end{tabular}
\end{center}
\caption{Survey Response Summary}
\label{table:table-2}
\end{flushleft}
\end{table}
Most responses indicate that human answers are perceived as more accurate, as measured by Question 1, which is indicated by Mode 4. Also, we can see the mode of agreement of question 5 is four, which indicates most of the respondents support human answers further for providing a better example. We try to understand the readability comparison between the human answer and the GPT-3 model's answer in question 2. Here, mode and median are both three, which means that readability-wise, the difference between a human answer and a ChatGPT answer is insignificant. On the other hand, in questions 3 and 4, we measure the conciseness and ease of understanding the answers. In these two metrics, experts ChatGPT answers less than Human answers, and in both cases, the mode of agreement was less than 3, and the median was 3.

\subsection{\textbf{Answer to RQ5:} Stack Overflow Traffic Trend Analysis}
\begin{figure}[ht]
\centering
\begin{subfigure}{.4\linewidth}
  \centering
  \includegraphics[width=\linewidth]{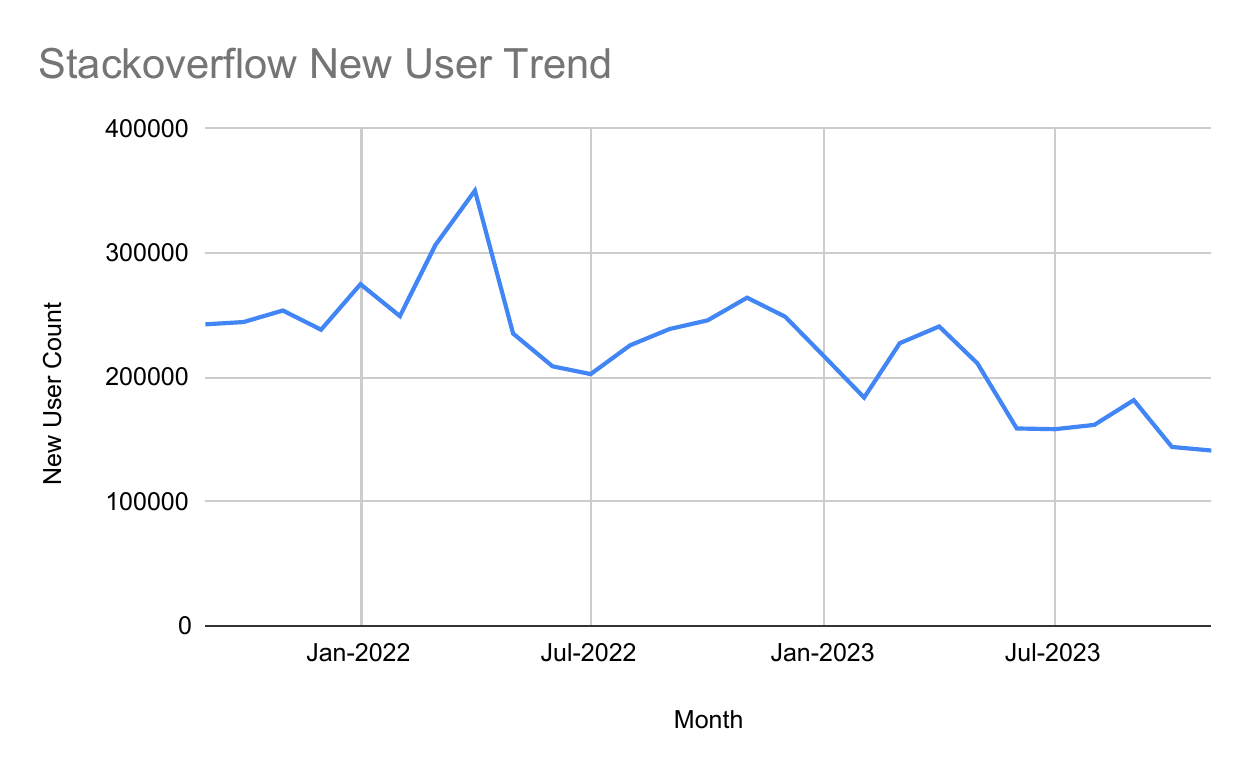}
  \caption{New User Trend}
  \label{fig:stacktraffic:sub1}
\end{subfigure}%
\hfill % This adds some space between the subfigures
\begin{subfigure}{.4\linewidth}
  \centering
  \includegraphics[width=\linewidth]{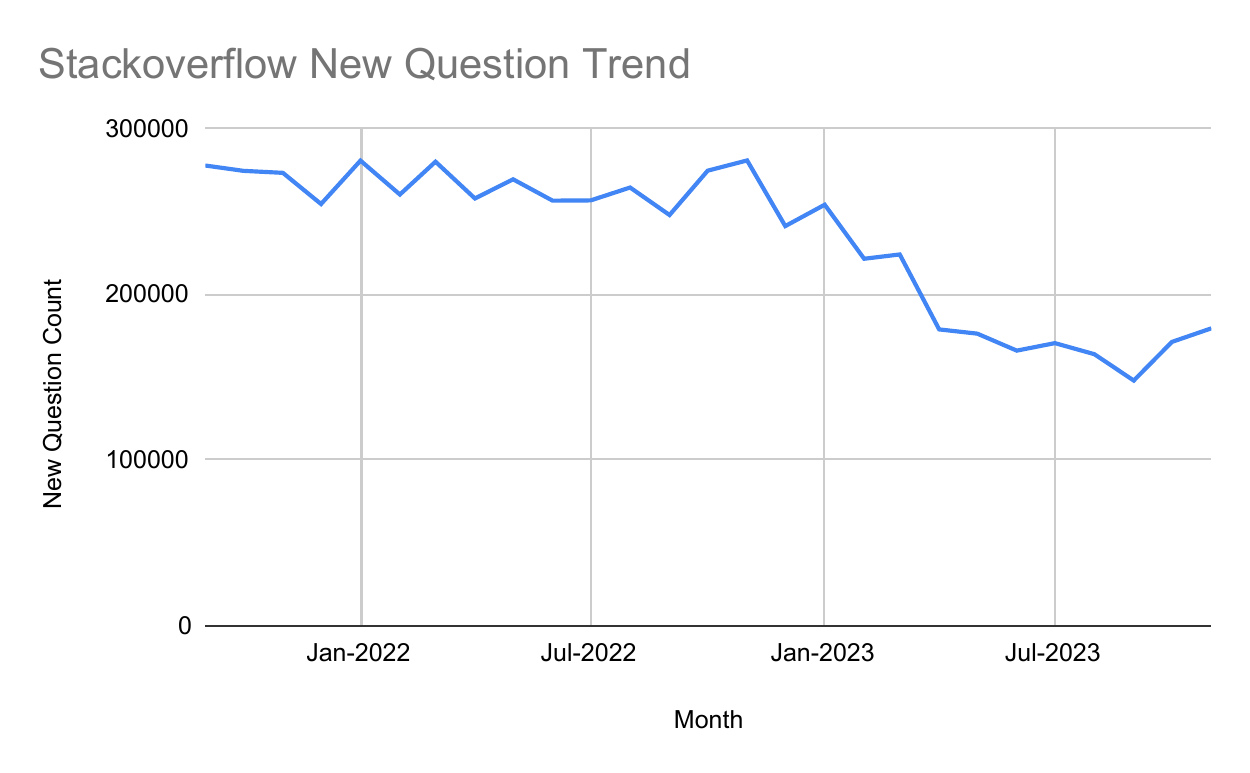}
  \caption{New Question Trend}
  \label{fig:stacktraffic:sub2}
\end{subfigure}
\hfill % This adds some space between the subfigures
\begin{subfigure}{.4\linewidth}
  \centering
  \includegraphics[width=\linewidth]{figs/Stackoverflow_New_Question_Trend-23-Dec.pdf}
  \caption{New Comment Trend}
  \label{fig:stacktraffic:sub3}
\end{subfigure}

\caption{Stack Overflow Traffic Trend of Last 2 years}
\label{fig:stacktraffic}
\end{figure}

We collect the monthly count of new registered users, new posts, and new comments and plot that in \ref{fig:stacktraffic}. We see some interesting patterns in the trend. For example, new user acquisition has always fluctuated since September 2021. Stack Overflow saw the biggest acquisition month in April 2022. ChatGPT was made available for users in November 2022, and immediately after that, there was no significant fall in user signup. However, in Feb-2023, we saw some decline, which again increased in Apr-2023 and the following month. Since June 2023, the monthly acquisition count has not crossed 200k.
On the other hand, the monthly new question count fluctuated between 240k and 280k from September 2021 to January 2023. Since Feb-2023, this trend has been going downward and has not crossed 180k since Apr-2023.
Similar to the new question, we see a downward trend in new comments from March 2023. However, the new question trend has some months where it increased from the previous month, whereas the new comment trend has no such uptick post-Mar-2023. This indicates that overall interaction in the platform might be reducing.

\section{Discussion}
In this section, we will discuss the implications of our observations and findings. 

\subsection{Better Linguistic Characteristics of GPT-3 Model}
Our numerical metrics analysis reveals that the responses generated by the GPT-3 model exhibit superior linguistic attributes. For instance, the GPT-3 model produces answers that are 66\% shorter than those generated by humans while simultaneously containing 35\% more words. From a cognitive perspective, ChatGPT's responses exhibit a 25\% higher degree of positivity. Nevertheless, the responses provided by the domain experts are somewhat inconsistent when considering the numerical aspect. Although 65\% of respondents expressed a favorable opinion regarding the linguistic features of ChatGPT, practitioners' opinions suggest minimal disparity in terms of readability and conciseness, as indicated by a rating of 3 for survey question 2. Given that all domain experts possess undergraduate degrees from universities, it is comprehensible that both responses may appear indistinguishable from them. 

\subsection{Lower Perceived Accuracy and Example of GPT-3 Model}
To avoid any positional bias in the answers, in one survey, we have put the human-provided answer in the first position, and the other survey, we put the ChatGPT answer in the first position. In terms of accuracy and example, 57\% of experts deemed that human-provided answers performed better. Though the practical accuracy of the answers provided by ChatGPT was between 70\% to 75\%, perceived accuracy seems a bit lower. The main reason behind that seemed to the the examples provided with the answers. ChatGPT often does not provide any example code unless explicitly prompted for. As accuracy and example are essential parts of this domain, we see scope for improvement for the LLMs here. During this study, more powerful models like GPT-4/5 were in the pipeline. In the future, we can measure whether those models improve accuracy and example.

\subsection{Programming Language Wise Behavior}
While comparing different metrics concerning programming languages, we observed some divergent behavior between JavaScript and Python. Specifically, we observed that the Python code length of the GPT-3 model's responses was 62\% shorter than the JavaScript code length. Examining the code similarity between the question and answer codes, we discovered that JavaScript answers were approximately 28\%  more similar to the question code than Python answers. The Automated Readability Index (ARI) and Flesch Reading Ease (FRE) metrics for the two languages did not differ significantly. In addition, our analysis of the polarity context revealed that Python-related GPT-3 model responses had a relatively higher proportion of positive sentiment. These findings highlight the significance of considering the programming language when evaluating the performance of language models for code generation tasks.

\subsection{Future of online CQA Platforms}
With the emergence of LLMs, the future of online CQA platforms will be increasingly exciting. The release of GPT-3 by ChatGPT, with its advanced user interface, has generated considerable interest among programmers. The influences are being studied and tracked extensively in many fields, including education \citep{lund2023chatting}, health care \citep{patel2023chatgpt}, public assessments \citep{gilson2023does}, and scholarly studies \citep{king2023conversation}. However, the effects of online CQA can be both beneficial and harmful.
Theoretically, we have demonstrated that ChatGPT answers are often better than human answers. We measured this from four different linguistics and cognitive metrics and the accuracy of the answers. From there, we have anticipated that people would be more engaged in ChatGPT-like tools for software engineering-related question answers, which may negatively impact the social interactions at Stack Overflow community-driven platforms. We finally provide data indicating that different interactions in those platforms have recently declined.
Platform traffic data collected from Stack Overflow also show a downtrend in parameters like new user creation, new questions, and new comments posted. Though ChatGPT was launched in Nov-2022, we have seen the downtrends from Feb-2023 onward. It might be logical that people need some time to catch up with new technology. Between Sep-2022 and Sep-2023, on a year-over-year basis, we can see a 40\% reduction in new question creation, a 24\% reduction in new user creation, and a 38\% reduction in the interactions in comments in Stack Overflow. However, historical data for the previous year shows that the traffic has some natural seasonality. So, it might be too early to decide whether a direct correlation exists between using LLM-driven models and the CQA downtrend.

\par In general, it is expected that people will prefer GPT-3 for its low waiting time. For example, for exhausted topics, the ETA of an answer to a question can be up to 400 hours \citep{ali2020modeling}, whereas ChatGPT gives those answers interactively. Also, interactive question answering, or chatting, has the benefit of asking quicker follow-up questions. Apart from that, people often face negative sentiments on social CQA platforms like Stack Overflow \citep{novielli2018gold}, which have a lower probability in a ChatGPT chat as our research exhibits a higher positive sentiment in ChatGPT's answer. Apart from that, as Chatgpt's answers are more concise and less verbose, that is the preferable way to deal with them for some practitioners. These factors can become important to drive more users away from Stack Overflow to ChatGPT.

Moreover, there are more butterfly affects of this. people increasingly interact with GPT-3 models for instant answers or to avoid uncertain adversity; there may be a gradual decrease in askers. It is well-established that sustaining the growth of social question-and-answer platforms requires retaining both askers and answerers \citep{kang2022motivational}. A decline in askers could lead to fewer answerers, as askers often become answerers in the social network chain \citep{bachschi2020asking}. Consequently, the knowledge network may become skewed

\par In hindsight, online CQA platforms have some recognition and reward mechanisms, which have a long-term social impact. One successful human answer has social and temporal factors \citep{calefato2015mining} apart from accuracy and linguistic factors. Different domain-specific community emergences and their evolutions have been pivotal in the history of these CQA platforms, and those have been studied in many research \citep{moutidis2021community, zagalsky2018r}. So, the diffusion of knowledge and the ever-growing knowledge network are also significant outcomes of using online CQA platforms. For example, previous studies indicated that the \textit{exchange of knowledge} and \textit{social interaction} have a positive impact on the community along with peer recognition and repudiation \citep{mustafa2023motivates} whereas while finding the answers from ChatGPT users will be missing the social interaction and peer recognition entirely. Hence the knowledge sharing graphs might evolve differently in future as an impact of that. People could grow more accustomed to receiving solutions in a personalized manner, at the expense of human interaction. Furthermore, the volume of publicly available questions and answers might decrease, potentially jeopardizing the openness and crowdsourcing of knowledge. Therefore, the increasing use of LLMs might affect the community structure, knowledge network, and information flow of online CQA platforms. Future innovations might hinge on how LLM models produce answers and content.

\section{Conclusion}
In conclusion, our findings suggest that GPT-3 models hold significant potential to revolutionize community-based question-answering platforms. They consistently outperformed human-provided answers regarding conciseness, readability, word sharing, understandability, and positivity, with a notable accuracy rate of 70-75\%. Despite a slight preference among domain experts for human responses, GPT-3 has coincided with a notable decline in new question creation, new user creation, and interactions in comments on Stack Overflow. Findings like these suggest that with users increasingly relying on AI-generated answers, GPT-3 improves the quality of online learning; it also challenges the status quo dynamics and relevance of community-based learning platforms. More research must be done to fully understand this change and find the best way to incorporate AI into community-based learning platforms.

\par \textit{Declaration of generative AI and AI-assisted technologies in the writing process}: During the preparation of this work the authors used Grammarly, ChatGPT, BARD and Quillbot in order to check their sentences for correct grammar and syntax. After using this tool/service, the authors reviewed and edited the content as needed and take full responsibility for the content of the publication.
\bibliography{cas-refs}
\bibliographystyle{model1-num-names}


% \bibliographystyle{apa}


% Author biography without author photo.
% Author biography. Author biography. Author biography.
% Author biography. Author biography. Author biography.
% Author biography. Author biography. Author biography.
% Author biography. Author biography. Author biography.
% Author biography. Author biography. Author biography.
% Author biography. Author biography. Author biography.
% Author biography. Author biography. Author biography.
% Author biography. Author biography. Author biography.
% Author biography. Author biography. Author biography.

% \bio{figs/pic1}
% Author biography with author photo.
% Author biography. Author biography. Author biography.
% Author biography. Author biography. Author biography.
% Author biography. Author biography. Author biography.
% Author biography. Author biography. Author biography.
% Author biography. Author biography. Author biography.
% Author biography. Author biography. Author biography.
% Author biography. Author biography. Author biography.
% Author biography. Author biography. Author biography.
% Author biography. Author biography. Author biography.

% \bio{figs/pic1}
% Author biography with author photo.
% Author biography. Author biography. Author biography.
% Author biography. Author biography. Author biography.
% Author biography. Author biography. Author biography.
% Author biography. Author biography. Author biography.

\end{document}